# Electromagnetically induced transparency at high optical power

**S. M. Iftiquar**
Department of Physics, Indian Institute of Science, Bangalore – 560012, India
Email: smiftiquar@gmail.com

*Abstract*

*We observe electromagnetically induced transparency (EIT) in Rb vapor at various optical intensities, starting from below saturation to several times the saturation intensity. The observed Lorentzian width of the EIT signal is very small. Solving the time dependant density matrix equation of motion with a phenomenological decay constant, we find an expression suitable in explaining the EIT signal. In this experimental observation and theoretical analysis intensity of EIT signal and its Lorentzian width increases with Rabi frequency of optical field.*

## INTRODUCTION

Recently a great deal of interest has been drawn on EIT because of its sub-natural spectroscopic feature. It also has potential application in quantum information processing, in which quantum nature of the EIT can be exploited [1]. In spectroscopy the natural decay rate broadens the spectral feature and creates uncertainty in reference frequency standard. For example, in laser locking by saturated absorption spectroscopy, the laser frequency remains uncertain by the decay constant of the excited state, for a 6 MHz decay constant we observe spectral line width of about 15 MHz. Thus, for more accurate laser frequency stability sub-natural spectral feature is an advantage. There are several techniques developed to create such narrow line. One of such techniques is to use a second laser connecting some other atomic transition that can create sharp a sharp transition [2]. The most interesting of these are EIT and coherent population trapping (CPT) techniques. In CPT the two laser fields are phase locked so that random phase noise between the fields are minimized. Additionally, in order to observe CPT spectra the vapor cell have to be filled up with buffer gas, lasers to be phase locked etc. In case of EIT the experimental system is simpler although the observed width is larger than that with CPT. Theoretically the analysis of EIT and CPT spectra should be different because for CPT the atoms interact with optical fields longer, whereas in case of EIT the atoms do not interact long enough.

EIT is a result of destructive quantum interference of absorption of two radiation fields in presence of atomic medium. Since the work of Harris [3,] a large number of experimental and theoretical investigation have been carried out on probe absorption and dispersion in EIT configuration. Interesting outcome of EIT is slowing [4] and trapping of light, quantum information processing [5] etc. It has been popularly known that coherent population trapping (CPT) and EIT are characteristically very similar to each other [6], although CPT condition requires intensity of the two fields to be similar while EIT can occur at any optical intensity. In 1995 Jyotsna et al. demonstrated that at high optical power strong CPT resonance takes place [7] while at lower optical intensity the same CPT resonance takes longer time to build up.

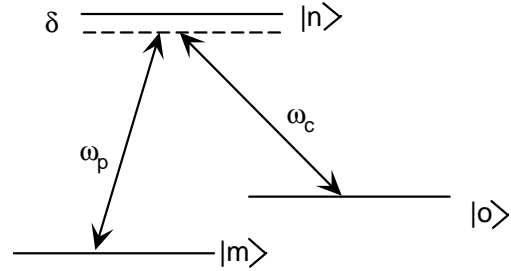

**Fig 1 :** Energy level diagram of 3-level lambda system for EIT.

## THEORY

In many of the theoretical analysis a density matrix equation of motion is taken and solved under the steady state condition, where time derivative of density matrix components are equated to zero. It is to be noted that such a steady state condition is not fully applicable in our experimental configurations because of finite beam width, low atomic number density etc. So we take individual atom photon interaction separately. Because of finite photon number density within the laser beam and dilute gas, finite beam width and thermal motion of atom, a finite number of atom photon interaction takes place with each atom and photon. Furthermore the density matrix $\rho$ concerned with the atomic transition is related to possible atomic states for individual atom. The Heisenberg-Langevin equation of motion for the density matrix can be written as:

$$\frac{\partial \rho}{\partial t} = -\frac{i}{\hbar}[H,\rho] - \rho\gamma$$

where $H$ is total Hamiltonian, $\gamma$ phenomenological decay constant of density matrix. The theoretical analysis can be simplified if semi classical picture is adopted in which atomic energy level is quantized but the optical field is considered classically. This is also true in dipole approximation where optical wavelength is much longer than atomic dimension. In this situation the dominating atom-field interaction will be of electric dipole type. In this situation atom field interaction Hamiltonian will be $H_I = -\vec{\mu}.\vec{E}$, where $\vec{\mu}$ is electric dipole moment vector of atom. Probe absorption in an atomic medium is associated to transition of the atom from lower state to higher energy state. This transition probability is directly related to respective off-diagonal term of the density matrix or coherence $\rho_{mn}$ for $m \neq n$. Through perturbative expansion the equation of motion of the coherence $\rho^{(1)}{}_{nm}$ can be expressed as

$$\frac{\partial \rho^{(1)}{}_{nm}}{\partial t} = -i\omega_{nm}\rho^{(1)}{}_{nm} - \frac{i}{\hbar}\left[H_I, \rho^{(0)}\right]_{nm} - \rho^{(1)}{}_{nm}(\gamma_{nm} + f)$$

where $\omega_{nm}$ transition frequency between atomic states $|n\rangle, |m\rangle$ and $\gamma_{nm}$ coherence decay rates at zero optical intensity, $f$ enhanced decay rate due to optical field, which is a function of Rabi frequency of optical field, $\Omega$. Solving the time dependant differential equation with the condition $\rho^{(0)}{}_{nm} = 0$, and taking $\rho^{(0)}{}_{mm}$ as population of state $|m\rangle$, and $\vec{E}(t) = E(\omega)e^{-i\omega t}$

$$\rho^{(1)}{}_{nm} = \frac{(\rho^{(0)}{}_{mm} - \rho^{(0)}{}_{nn})}{\hbar} \frac{\vec{\mu}_{nm} \cdot \vec{E}(t)}{(\omega_{nm} - \omega) - i(\gamma_{nm} + f)},$$

so the polarization of the medium induced by $\vec{E}(t)$, can be expressed as

$$P_{ij}^{(1)}(\omega) = \sum_{m,n} \frac{N(\rho^{(0)}{}_{mm} - \rho^{(0)}{}_{nn})}{\hbar} \frac{\mu^i_{nm}\mu^j_{mn}E_i(t)}{(\omega_{nm} - \omega) - i(\gamma_{nm} + f)}$$

Where the summation is over near-degenerate states and i, j corresponds to direction of incident laser polarization and direction of observation. Absorption of the field $\alpha(\omega)$ can be defined as $\alpha(\omega) = \frac{4\pi \, \text{Im}(P^{(1)}{}_{ij})}{c}$ where c is speed of light.

In the EIT experiment two laser fields couple in the atomic medium so, due to optical pumping effect of the control laser, the probe absorption can be expressed as

$$Abs(\omega_p) = P_i(\omega_p)[1 + GP_k(\omega_c)]$$

where $\omega_p, \omega_c$ are probe and control laser frequency respectively, k is polarization direction of control laser, G is a factor that determines optical pumping effect of control laser on probe absorption, $G \leq 1$.

**EXPERIMENTS**

The experimental setup is shown in figure 2. We used [87]Rb isotope for the experiment. The lower ground state $|m\rangle$ corresponds to ground state hyperfine level $F = 1$ and $|n\rangle \equiv F' = 1$ and $|o\rangle \equiv F = 2$. Control laser is frequency locked with the help of a saturated absorption spectroscopy in a counter propagating pump probe configuration (Rb3) and the probe is frequency scanned with help of piezo-electric actuator. The probe spectra is recorded with the help of Rb cell Rb2 and photo-diode PD2. The control laser and probe laser power towards the EIT cell is controlled with the help of half-wave plate and polarizing cube beam splitter. The magnetic shield is used to cut off stray magnetic field that may be present around the EIT cell. The Rb cells length is 5 cm and diameter 2.5 cm, while the laser beam diameter is 2 mm, obtained by an aperture (A) placed before the EIT cell. EIT spectra can be obtained under various conditions, however in the following, the spectra has been obtained when control laser is locked to $F = 2 \rightarrow F' = 1$ transition and the probe laser is slowly frequency scanned over $F = 1 \rightarrow F' = all$ frequency band. The EIT spectra is obtained when probe is at resonance to $F = 1 \rightarrow F' = 1$ transition frequency. The Rb cell contains [85]Rb 72.2% and [87]Rb 27.8%, which is natural abundance.

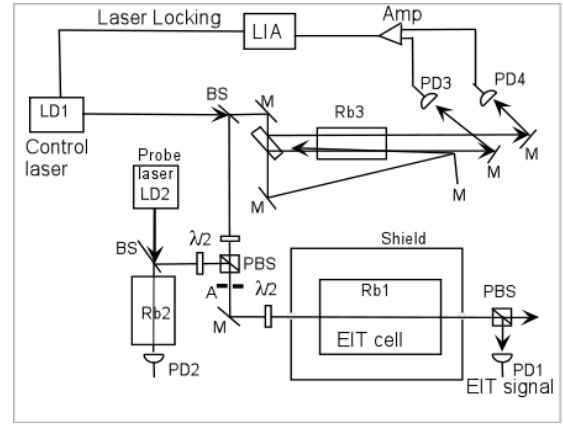

**Fig 2 :** Experimental setup. LD diode laser, LIA lock in amplifier, PD photo-diode, Rb rubidium cell, PBS polarizing cube beam splitter, BS beam splitter, $\lambda/2$ half-wave plate, 'shield' magnetic shield, A aperture.

**RESULTS AND DISCUSSIONS**

Figure 3 shows EIT spectra at various optical intensity. Vertical shift of the traces are arbitrary, in order to show the traces distinctly. The sharp transparency dip is due to electromagnetically induced transparency resonance at the center of absorption for $F = 1 \rightarrow F' = 1$ transition, which can also be called hyperfine absorption. Intensity of hyperfine absorption increases with increased control laser intensity. It can also be noted that intensity and width of EIT spectra also increases with $\Omega_c$.

A theoretical simulation to the spectra can also be drawn based on the theory developed here, figure 4 shows the traces of EIT spectra for various control laser Rabi frequency $\Omega_c$. Comparing the theoretical plot to the experimental results, it is clear that the theory can explain the EIT spectra very successfully. The main difference with the conventional theory [2] is that in this analysis we have considered the dynamic solution of the density matrix equation in a perturbative manner, that implies steady state interaction between atom and photon is not achieved before atom and /or photon moves out of the interaction region.

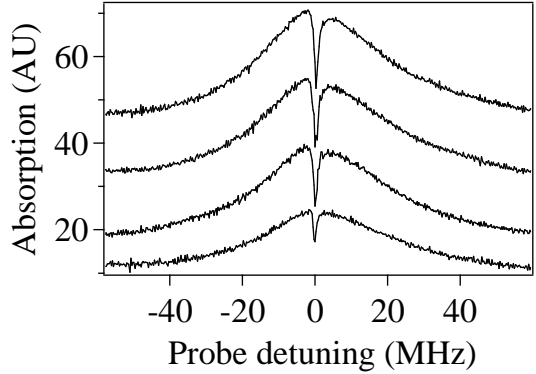

**Fig 3 :** Experimentally obtained EIT spectra for probe at $^{87}Rb$; $F = 1 \rightarrow F' = 1$ with Rabi frequency $\Omega_p \approx 6$ MHz, and scanning frequency to cover all excited state hyperfine states. $\Omega_c = 7; 6.2;$ 5.4, 3.4 MHz from top to bottom trace.

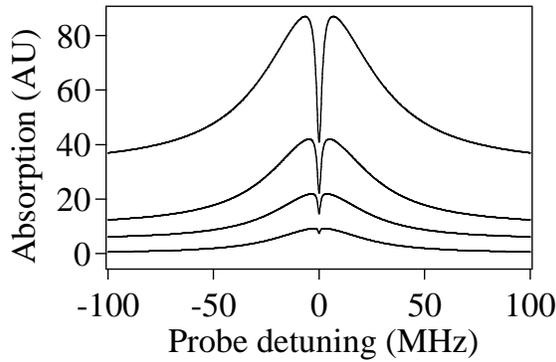

**Fig 4 :** Theoretical trace of EIT spectra, upper trace for control laser Rabi frequency $\Omega_c = 52$ MHz, and rest $\Omega_c$ - = 22, 8, 2 MHz respectively.

Figure 5 shows Lorentzian half width of a collection of several EIT spectra. The straight line is drawn as a guide to the eye, which indicates that the variation of EIT width with control laser Rabi frequency is approximately linear, with $f(\Omega_c) = 0.486 + 0.025\Omega_c$. This means, extrapolating this line to $\Omega_c = 0$ we get combined laser line width reflected on the EIT spectra as 486 KHz, or assuming each laser of similar type and there are no other EIT width broadening mechanism, the laser line width can be estimated as $\Delta = 486/\sqrt{2} = 344$ KHz. It should be noted that there are several factors that influence line width broadening of the EIT spectra. For example incoherent collision with other atoms and wall of the container. This can be reduced by using high pressure buffer gas and paraffin coating on the cell wall. Another broadening mechanism is due the random thermal speed of atoms. As atom crosses laser beam path it introduces transit time broadening to the spectra. It can be expressed as $\gamma_t = \sqrt{\frac{2k_B T}{\pi D^2 m}}$ where $k_B$ Boltzmann constant, T cell temperature = 300 K, D laser beam diameter, m Rb atomic mass. The estimated value for $\gamma_t$ = 68 KHz. This can be reduced by using larger beam diameter and lower gas temperature.

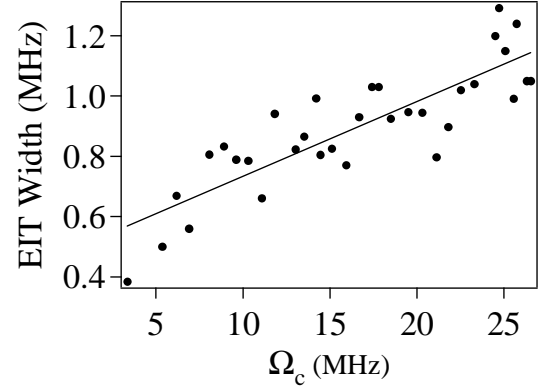

**Fig 5:** Variation of Lorentzian width of EIT spectra at various optical power.

## CONCLUSIONS

A very sharp sub-natural line width EIT signal has been obtained with ordinary Rb vapor cell. It is possible to estimate laser line width with through this EIT technique. The narrow feature is further useful in high resolution spectroscopic measurements, slowing down speed of light and quantum information processing.

## ACKNOWLEDGEMENTS

In Acknowledgment, this work was supported by the Council of Scientific and Industrial Research of India.